\documentclass[11pt,a4paper]{article}
\usepackage{jheppub}
\usepackage{graphicx,epsf}
\usepackage{amssymb}
\usepackage{color}

\def\SU2{\ensuremath{{SU(2)}}}

\newcommand{\bea}{\begin{eqnarray}}
\newcommand{\eea}{\end{eqnarray}}

\def\beq{\begin{equation}}
\def\eeq{\end{equation}}
\def\beqn{\begin{eqnarray}}
\def\eeqn{\end{eqnarray}}

 \title{Emission spectra of self-dual black holes}
\author[a]{Sabine Hossenfelder}
\author[b]{ Leonardo Modesto}
\author[c]{Isabeau Pr\'emont-Schwarz}
 
\affiliation[a]{NORDITA, Roslagstullsbacken 23, 106 91 Stockholm, Sweden }

\affiliation[b]{Perimeter Institute for Theoretical Physics, 31 Caroline St.N., Waterloo, ON N2L 2Y5, Canada }
\affiliation[c]{Max-Planck-Institut f\"ur Gravitationsphysik, Am M\"uhlenberg 1, Golm, D-14476, Germany }
\emailAdd{lmodesto@perimeterinstitute.ca}

\abstract{

We calculate the particle spectra of evaporating self-dual black holes that are potential dark matter
candidates. We first estimate the relevant mass and temperature range and find that the masses are
below the Planck mass, and the temperature of the black holes is small compared to their mass. 
In this limit, we then derive the number-density of the primary emission particles, and, by 
studying the wave-equation of a scalar field in the background metric of the black hole, show that we 
can use the low energy approximation for the greybody factors. We finally arrive at the 
expression for the spectrum of secondary particle emission from a dark matter halo constituted
of self-dual black holes.
}
 
\keywords{Black holes, black hole evaporation, quantum gravity} 

\begin{document}
\maketitle

\section{Introduction}

Observables of quantum gravity are hard to come by. Compared to the three other interactions of the Standard Model,
gravity is extremely weak, and one expects quantum effects of gravity to
become relevant only when the space-time curvature reaches the Planckian regime.
Unfortunately, the enormously high energy densities necessary to
achieve sufficiently strong curvature to make for observable effects are inaccessible in the
laboratory\footnote{Unless
gravity is only seemingly weak and the true Planck energy can be reached
at particle colliders as in
scenarios with large additional dimensions. We will not
consider this possibility here.}.

There
are then two ways one can hope to find signatures arising from strong
curvature effects. The one is focusing on the early universe, the other one
is black holes. In both cases the singularity theorems tell us if
classical General Relativity remained valid then the curvature would not
only reach the Planckian regime, but would diverge.
Such instances of infinite densities seem unphysical and one expects
effects of quantum gravity to prevent the formation of singularities.

One approach to quantum gravity, Loop Quantum Gravity
(LQG) \cite{LQGgeneral,LQGgeneral2,LQGgeneral3,Mercuri:2010xz}, has given rise to
models that allow to describe the very early universe. Simplified
frameworks of LQG using a minisuperspace approximation has been shown
to resolve the initial singularity problem \cite{Bojowald:2006da,MAT}.
In the present work we will study the properties of black
holes in such a minisuperspace model. The metric of black
holes in this model was previously derived in \cite{Modesto:2008im},
where it was shown in particular that the singularity is removed by
 a self-duality of the metric that replaces the black hole's
usually singular inside by another asymptotically flat region.  The
thermodynamical properties of these self-dual black holes have been
examined in \cite{Modesto:2009ve,Alesci:2011wn}, and
in \cite{Hossenfelder:2009fc} the dynamical aspects of the collapse and
evaporation were studied. 

Here, we want to work towards the phenomenology of the self-dual black
holes. The previous analysis has shown that the late stage of the evaporation
process is significantly modified by quantum gravitational effects. The
temperature of the black holes is lowered and they form quasi-stable
remnants with an infinite evaporation time. If these objects were
formed primordially, they would constitute dark
matter candidates and their slowly proceeding particle emission
could become observable today. We will thus here study the evaporation
process, and provide the theoretical prerequisites necessary to make
contact with observation.

This paper is organized as follows. In the next we will
section briefly summarize
the previously derived self-dual black hole metric and recall its thermodynamical
properties.
In section \ref{thermo} we study the emission and the greybody-factors and arrive
at the final expression for the emission spectrum. We
discuss the next steps that are to be taken to make contact with observation,
and summarize in section \ref{further}.

In what follows, we will use Planck units.

\section{The metric of the self-dual black hole}
\label{metric}

The regular black hole metric that we will be using is derived from a simplified model of {\sc LQG} \cite{Modesto:2008im}.
{\sc LQG} is based on a canonical quantization of the Einstein equations written in terms of the Ashtekar variables \cite{AA}, that is in terms of an $\mathrm{su}(2)$ 3-dimensional connection $A$ and a triad $E$. The basis states of {\sc LQG} then are closed graphs the edges of which are labeled by irreducible $\mathrm{su}(2)$ representations and the vertices by $\mathrm{su}(2)$ intertwiners (for a review see e.g. \cite{Mercuri:2010xz}). The
edges of the graph represent quanta of area with area $\gamma l_{\rm P}^2 \sqrt{j(j+1)}$, where $j$ is a half-integer representation label on the edge, $l_{\rm p}$ is the Planck length,  and $\gamma$ is a parameter of order 1 called the Immirzi parameter. The vertices of the graph represent quanta of 3-volume. One important consequence that we will use in the following is that the area is quantized and the smallest possible quanta correspond to an area of$\sqrt{3}/2 \gamma l_{\rm p}^2$.

To obtain the simplified black hole model the following assumptions were made. First, the number of variables was reduced by assuming spherical symmetry. Then, instead of all possible closed graphs, a regular lattice with edge lengths $\delta_1$ and $\delta_2$ was used. The solution was then obtained dynamically inside the homogeneous region (that is inside the horizon where space is homogeneous but not static). An analytic continuation to the outside of the horizon shows that one can reduce the two free parameters by identifying the minimum area present in the solution with the minimum area of {\sc LQG}. The one remaining unknown constant $\delta$ is a parameter of the model determining the strength of deviations from the classical theory, and would have to be constrained by experiment. With the plausible expectation that the quantum gravitational corrections become relevant only when the curvature is in the Planckian regime, corresponding to $\delta < 1$, outside the horizon the solution is the Schwarzschild solution up to negligible Planck-scale corrections.

This quantum gravitationally corrected
Schwarzschild metric can be expressed in the form
\begin{eqnarray}
ds^2 = - G(r) dt^2 + \frac{dr^2}{F(r)} + H(r) d\Omega~,
\label{g}
\end{eqnarray}
with $d \Omega = d \theta^2 + \sin^2 \theta d \phi^2$ and
\begin{eqnarray}
&& G(r) = \frac{(r-r_+)(r-r_-)(r+ r_{*})^2}{r^4 +a_0^2}~ , \nonumber \\
&& F(r) = \frac{(r-r_+)(r-r_-) r^4}{(r+ r_{*})^2 (r^4 +a_0^2)} ~, \nonumber \\
&& H(r) = r^2 + \frac{a_0^2}{r^2}~ .
\label{statgmunu}
\end{eqnarray}
Here, $r_+ = 2m$ and $r_-= 2 m P^2$ are the two horizons, and $r_* = \sqrt{r_+ r_-} = 2mP$. $P$ is the
polymeric function $P = (\sqrt{1+\epsilon^2} -1)/(\sqrt{1+\epsilon^2} +1)$, with
$\epsilon \ll 1$ the product of the Immirzi parameter ($\gamma$) and the polymeric parameter ($\delta$): $\epsilon = \delta \gamma$. Under these conditions, we
also have $P \ll 1$, and so $r_-$ and $r_*$ are very close to $r=0$. The area $a_0$ is equal to $A_{\rm min}/8 \pi$, $A_{\rm min}$ being the minimum area 
of {\sc LQG}, which we will assume to be  
of the order Planck length squared.

Note that in the above metric, $r$ is only asymptotically the usual radial
coordinate since $g_{\theta \theta}$ is not just $r^2$. This choice of
coordinates however has the advantage of easily revealing the properties
of this metric as we will see. But first, most importantly, in the limit
$r \to \infty$ the deviations from the Schwarzschild-solution are of
order $M \epsilon^2/r$, where $M$ is the usual {\sc ADM}-mass:
\beqn
G(r) &\to& 1-\frac{2 M}{r} (1 - \epsilon^2)~, \nonumber  \\
F(r) &\to& 1-\frac{2 M}{r}~ , \nonumber \\
H(r) &\to& r^2 .
\eeqn
The {\sc ADM} mass is the mass inferred by an observer at flat asymptotic infinity; it is determined solely
by the metric at asymptotic infinity.  The parameter $m$ in the solution is related to the mass $M$ by $M = m (1+P)^2$.

If one now makes the coordinate transformation $R = a_0/r$ with the rescaling
$\tilde t= t \, r_*^{2}/a_0$, and
simultaneously substitutes $R_\pm = a_0/r_\mp$, $R_* = a_0/r_*$ one finds that the metric in
the new coordinates has the same form as in the old coordinates and thus exhibits a
very compelling type of self-duality with dual radius $r=\sqrt{a_0}$. Looking at the angular part
of the metric, one sees that this dual radius corresponds to a minimal possible
surface element. It is then also clear that in the limit $r\to 0$, corresponding
to $R\to \infty$, the solution
does not have a singularity, but instead has another asymptotically flat Schwarzschild region.

The metric in Eq. (\ref{statgmunu}) is a solution of a quantum gravitationally corrected set of
equations which, in the absence of quantum corrections $\epsilon, a_0 \to 0$, reproduce Einstein's field equations.
However, due to these quantum corrections, the
above metric is no longer a vacuum-solution to Einstein's field
equations. Instead, if one computes the Einstein-tensor and sets it
equal to a source term $G_{\mu \nu} = 8 \pi \widetilde{T}_{\mu \nu}$, one
obtains an effective quantum gravitational stress-energy-tensor $\widetilde{T}_{\mu \nu}$,
which violates the positive energy condition. Since the positive energy condition is one of the assumptions
for the singularity theorems, this explains how the solution can be entirely regular.

\begin{figure}
 \begin{center}
 \includegraphics[height=8.5cm]{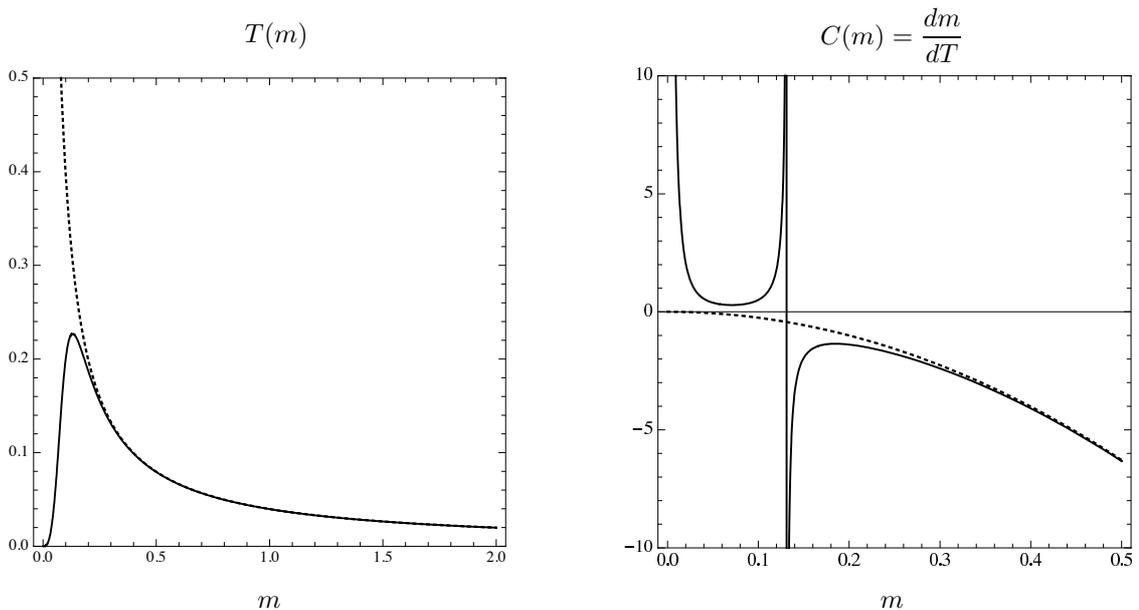}
      \end{center}
  \caption{\label{Potplot}
 Plot of the temperature $T(m)$ on the left and of the heat capacity $C_s =\frac{{\rm d} m}{{\rm d}T}$ on the right in Planck units.
The continuous lines represent the
quantities of the self-dual black hole, and
 the dashed lines represent the classical quantities. }
\label{temperature}
  \end{figure}

The derivation of the black hole's thermodynamical properties from this metric
is now straightforward and proceeds in the usual way.
 The Bekenstein-Hawking temperature $T_{BH}$ is given in terms of the surface gravity
$\kappa$ by $T_{BH} = \kappa/2 \pi$, and
\begin{eqnarray}
\kappa^2 = - g^{\mu \nu} g_{\rho \sigma} \nabla_{\mu} \chi^{\rho} \nabla_{\nu}
\chi^{\sigma}/2 = - g^{\mu \nu} g_{\rho \sigma}  \Gamma^{\rho}_{\;\mu 0} \Gamma^{\sigma}_{\;\nu 0}/2~,
\end{eqnarray}
where $\chi^{\mu}=(1,0,0,0)$ is a timelike Killing vector and $\Gamma^{\mu}_{\; \nu \rho}$
are the connection coefficients. Plugging in the metric, one obtain
\begin{eqnarray}
T_{BH}(m) = \frac{(2m)^3 (1 - P^2)}{4 \pi [ (2 m)^4 + a_0^2]} ~.
\label{Temperatura}
\end{eqnarray}
This temperature coincides with the Hawking temperature in the limit of
 large masses but goes to zero for $m \rightarrow 0$. We remind the reader that the black hole's {\sc ADM} mass $M=m (1+P)^2 \approx m$,
since $P \ll 1$. Fig.\ref{temperature} shows the
temperature as a function of
the black hole mass $m$.
We see that for small values of the mass there is a substantial difference between the
usual semiclassical temperature (dashed line) and the quantum gravitationally corrected temperature (continuous line).
In fact the semiclassical temperature tends to zero and does not diverge for $m\rightarrow 0$.
The temperature is maximum for $m^* = 3^{1/4} \sqrt{A_{\rm min}}/\sqrt{32 \pi}$
and $T^*= 3^{3/4} (1-P^2)/\sqrt{32 \pi A_{\rm min}}$.
It is noteworthy that $m^*$ depends only on the Planck area $A_{\rm min}$. 

From the temperature, one obtains the black hole's entropy by making use of the
thermodynamical relation $S_{BH}=\int dm/T(m)$.
Calculating this integral yields
\begin{eqnarray}
S= \frac{( 1024 \pi^2 m^4 - A_{\rm min}^2)(1+ P)^2}{256 \pi m^2 (1 - P^2)} + {\rm const.}.
\label{entropym}
\end{eqnarray}
We can express the entropy in terms of the event horizon area.
\begin{eqnarray}
A = \int {\rm d} \phi {\rm d} \theta \sin \theta \, p_c(r)\Big|_{r = 2m} = 16 \pi m^2 +  \frac{A_{\rm Min}^2}{64 \pi m^2}.
\label{area}
\end{eqnarray}
Inverting (\ref{area})
for $m=m(A)$ and inserting into (\ref{entropym})
we obtain
\begin{eqnarray}
 S = \pm \frac{\sqrt{A^2 - A_{\rm Min}^2}}{4} \, \frac{(1+P)}{(1-P)} ~ ,
\label{entropyarea}
\end{eqnarray}
where we have set the possible additional constant to zero. $S$ is positive for $m>\sqrt{a_0}/2$,
and negative otherwise.

\section{Emission}
\label{thermo}

From the black hole temperature one now commonly continues to calculate the evaporation
rate ${\rm d}M/{\rm d}t$ of the black hole by making use of Stefan-Boltzmann law such that
\begin{eqnarray}
\frac{{\rm d}M}{{\rm d}t} = (1+P)^2 \frac{{\rm d}m}{{\rm d}t} = \alpha A(m) T_{BH}^4(m),
\end{eqnarray}
where (for a single massless field with 2 degree of freedom)
$\alpha = \pi^2/60$ and $A(m)$ is the area of the event horizon. If one does
so, one sees that due to the drop of the temperature, the black hole's lifetime is
infinite, as has been shown in \cite{Modesto:2009ve}.

The purpose of this paper is to arrive at a more exact expression for the emission
rate that will be more suitable to make contact to experiment. 

For this we first notice that 1.) Even in the semi-classical case the use of
Stefan-Boltzmann's law is inappropriate when the typical energy of the
emitted particles becomes comparable to the total mass of the black hole.
One can then no longer treat the black hole as a heat bath and use the
macro-canonical ensemble, but one has to use the micro-canonical ensemble
taking into account the decrease in entropy caused by the emission of the
particle. Even without quantum gravitational effects,
this suffices to correct the unphysical divergence of the black hole's
temperature \cite{Casadio:2001dc,Casadio:2000py,Hossenfelder:2004af}. Then, 2.) we need to know the greybody
factors of the black hole caused by backscattering on the gravitational
potential that lead to deviations from the blackbody radiation. Next, 3.)
we have to take into account all elementary particle species with their
individual degrees of freedom and spin statistics and, finally, 4.)
we would have to integrate over
the fragmentation functions to obtain the spectrum of the outgoing
particles. 

We will in the following subsections address the first three points.
We will in this paper not present a full numerical study, but
provide the analytical expressions necessary for such a study. The
goal of our work is to make an important step towards examining the
viability of these quasi-stable self-dual black holes as dark matter
candidates. The emission spectra are the central ingredient to
identify them. 

Before we look into the details, let us recall which parameter ranges we are 
interested in, so that we can make suitable approximations to simplify our analysis. 
In \cite{Modesto:2009ve} it was estimated that primordial production of the
self-dual black
holes would be relevant only for masses smaller than $10^{-3} m_{\rm p}$.
For such small black hole masses, we have from Eq. (\ref{Temperatura})
that $T \approx m^3/m_{\rm p}^2$. As noted earlier, in contrast to the
normal case, the self-dual black holes get cooler
the smaller their mass.

We have to keep in mind here that since the
black hole's mass is below the Planck mass, the radius of the outer
horizon is inside the dual radius which is a minimum radius. This means
that the black hole's radiation has to pass through a `pinhole' of
Planck size. This case is, not coincidentally, very similar
to earlier considered `bag of gold' scenarios in which a (potentially
infinitely) large volume is contained inside a small surface area 
\cite{bog,Hsu:2007dr,Hossenfelder:2009xq}. As a consequence, the surface from which we can receive
radiation from the black hole is actually not the horizon area, but the minimal
area $\approx l_{\rm p}^2$. A rough estimate for the mass loss rate is then
\beqn
\frac{{\rm d}M}{{\rm d}t} \approx l_{\rm p}^2 T^4 \approx \frac{M^{12}}{m_{\rm p}^{10}} ~.
\eeqn
Integrating the inverse of ${\rm d}M/{\rm d}t$ to obtain the lifetime, one finds that the
time it takes for the black hole to completely evaporate exceeds the
lifetime of the universe for $m \gtrsim 10^{-5} m_{\rm p}$. The primordially produced
black holes with masses of about $10^{-3} m_{\rm p}$ thus would still not
have entirely decayed today. Moreover, they would have an average temperature
of $T \approx 10^{-9}m_{\rm p} \approx 10^{9}$~TeV, which is about in the
energy range of the ultra high energetic cosmic rays ({\sc{UHECR}}s) whose origin is still
unclear. We thus see why the self-dual black holes can make for an
interesting phenomenology. However, to arrive at observational consequences
we have to make this rough estimate more precise. For this, we take with us that
the parameter range we are interested in is $T \ll M \approx m \ll m_{\rm p}$.

The evaporation of regular Planck scale black holes has recently attracted a lot of attention,
and the emission properties of other types of regular black holes than the ones discussed here have been
considered in \cite{Leosfriends1,Leosfriends2,Leosfriends3,Leosfriends4,Leosfriends5,Leosfriends6}. 

\subsection{Statistics}

In the micro-canonical picture, we have that the number particle density for a
single particle with energy $\omega$ emitted from a black hole of mass $m$ 
(we recall the {\sc ADM} mass is $M = m (1+P)^2 \approx m$, in the rest of the paper we 
will refer to $m$ as the 
black hole mass) is 
\begin{eqnarray}
n_s (\omega) = e^{S_{BH}(m -\omega)-S_{BH}(m)} , \label{nomeg}
\end{eqnarray}
where $S_{BH}(m)$ is the entropy of a black hole of mass $m$. For bosons which can have
multiple particles in the same quantum state, the multiparticle number particle density is then:
\begin{eqnarray}
n_m (\omega) = \sum_{j=1}^{\lfloor m/\omega \rfloor} j e^{S_{BH}( m - j \omega) -S_{BH}(m)} ,\label{nomegs}
\end{eqnarray}
where $\lfloor \cdot \rfloor$ denotes the next smaller integer.
The self-dual black hole of mass $m$ has an entropy of
\begin{eqnarray}
S_{BH}(m) = \left(4 m^2-\frac{a_0^2}{4 m^2} \right) \pi . \label{entropii}
\end{eqnarray}
We wish to see when the macrocanonical approximation of
\begin{eqnarray}
n_M(\omega) = [e^{\omega/T(m)}\pm 1]^{-1}\label{macro}
\end{eqnarray}
breaks down. Taking the ratio of Eq(\ref{nomeg}) over (\ref{macro}), using (\ref{Temperatura}) with
$1-P^2 \approx 1$
\begin{eqnarray}
T(m) = \frac{2 m^3}{\left(a_0^2+16 m^4\right) \pi } ,\label{tempm}
\end{eqnarray}
and we find
\begin{eqnarray}
r(\omega) \equiv \frac{n_s(\omega)}{n_M(\omega)}\approx \exp{\left( \left[ -\frac{a_0^2}{4 m^2 (m-\omega )^2}-\frac{a_0^2}{2 m^3 (m-\omega )}+4\right] \pi  \omega ^2 \right) }\label{ratio}
\end{eqnarray}
for $\omega/T(m) \gg 1$. From Eq. (\ref{ratio}) we see that for $m\gg \sqrt{a_0}$, the difference between the macrocanonical and microcanonical starts to be significant for $\omega$ of the order of the Planck energy. For $m\ll\sqrt{a_0}$ however, the macro and microcanonical analysis begin to diverge when
\begin{eqnarray}
\frac{\omega^2 a_0^2}{ m^2 (m-\omega )^2}> 1 \Leftrightarrow \omega >m^2/a_0 \label{msq} ~.
\end{eqnarray}
When $m\ll \sqrt{a_0}$, it is therefore important to use the microcanonical analysis when, 
in Planck units, energies reach the level of $m^2/a_0$. In that case we then have
\begin{eqnarray}
n_s(\omega) &=& \exp \left( -\frac{\pi  \left(a_0^2+16 m^2 (m-\omega )^2\right) (2 m-\omega ) \omega }{4 m^2 (m-\omega )^2} \right) \nonumber \\
&\approx& \exp\left(-\frac{  \left(a_0^2 \right)  \omega }{ m (m-\omega )^2}  \right) \ll 1 	~.\label{limitz}
\end{eqnarray}
This further implies that
\begin{eqnarray}
n_s(2 \omega) \approx n_s(\omega)^2 \ll n_s(\omega)~.
\end{eqnarray}
In fact in such a case we have that
\begin{eqnarray}
\frac{\left|n_m( \omega) -n_s(\omega)\right|}{n_s({\omega})} < \frac{2 n_s(\omega) }{(1- n_s(\omega))^2}~,
\end{eqnarray}
so we can safely approximate
\begin{eqnarray}
n_s(\omega)\approx n_m(\omega)~.
\end{eqnarray}
We will use this approximation in the following.

\subsection{Greybody factors}
\label{greybody}

We will now calculate the propagation of a scalar field in the black hole's background to
analyze the backscattering on the potential well, which will in general depend on the angular momentum
of the field's modes. Our metric depends on the three functions $F(r)$, $G(r)$ and $H(r)$
defined in Eqs. (\ref{statgmunu}). It is always possible to introduce a new radial coordinate
$\tilde r$ such that $H(\tilde r) = \tilde r^2$. However, in this coordinate system the
metric coefficients become quite complicated expressions that are in addition
only piecewise defined. We will thus continue to use the form of the metric introduced in the first
section, but have to keep in mind that the coordinate $r$ agrees only asymptotically with
the usual radial coordinate while for small $r$ it bounces on the self-dual radius corresponding
to the minimal possible area. We follow here the usual procedure that can be found for example in 
\cite{Perturbations, MacGibbon:1991tj}.

\begin{figure}
 \begin{center}
 \hspace{-0.4cm}
  \includegraphics[width=6.6cm]{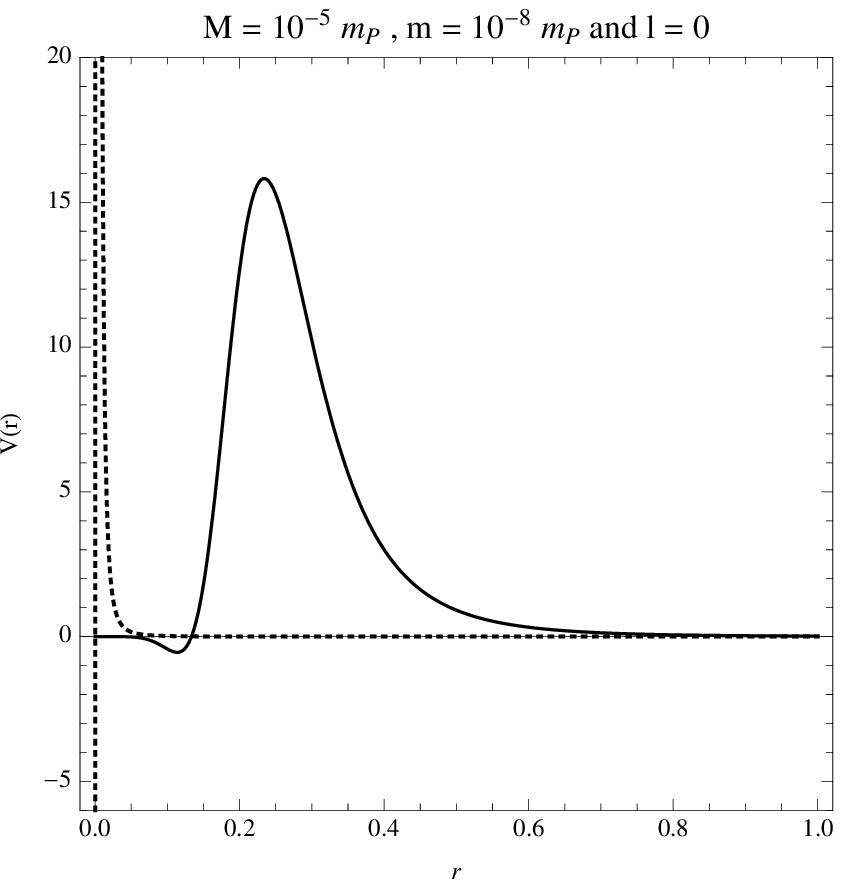} \hspace{0.2cm}
  \includegraphics[width=6.6cm]{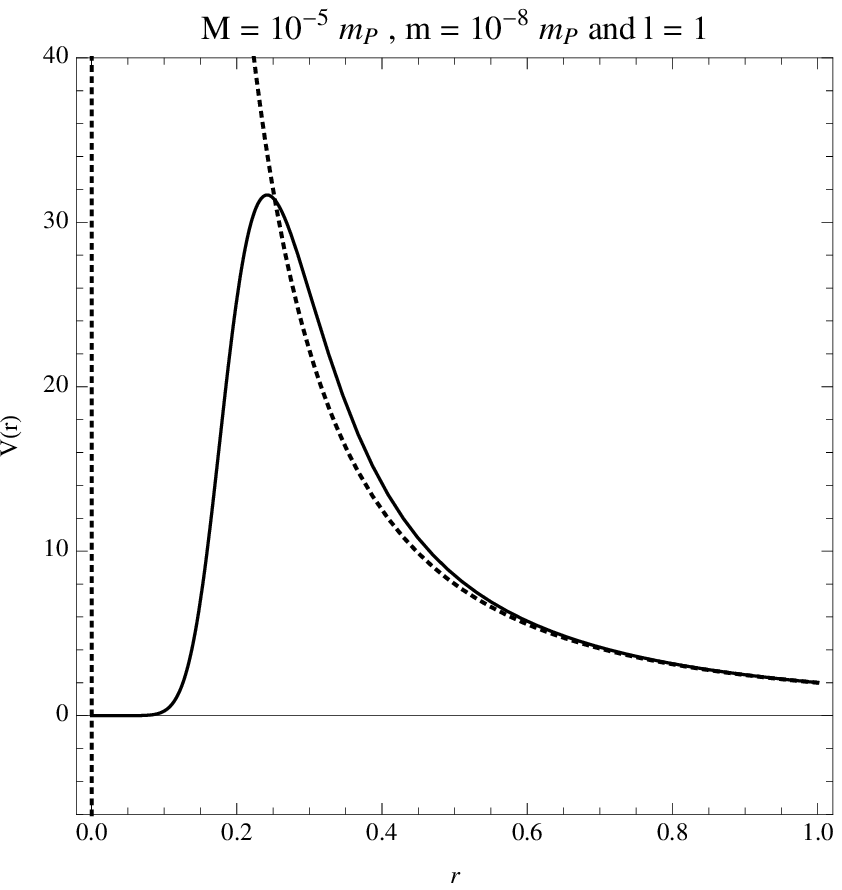}
\hspace{0.0cm}
  \includegraphics[width=6.6cm]{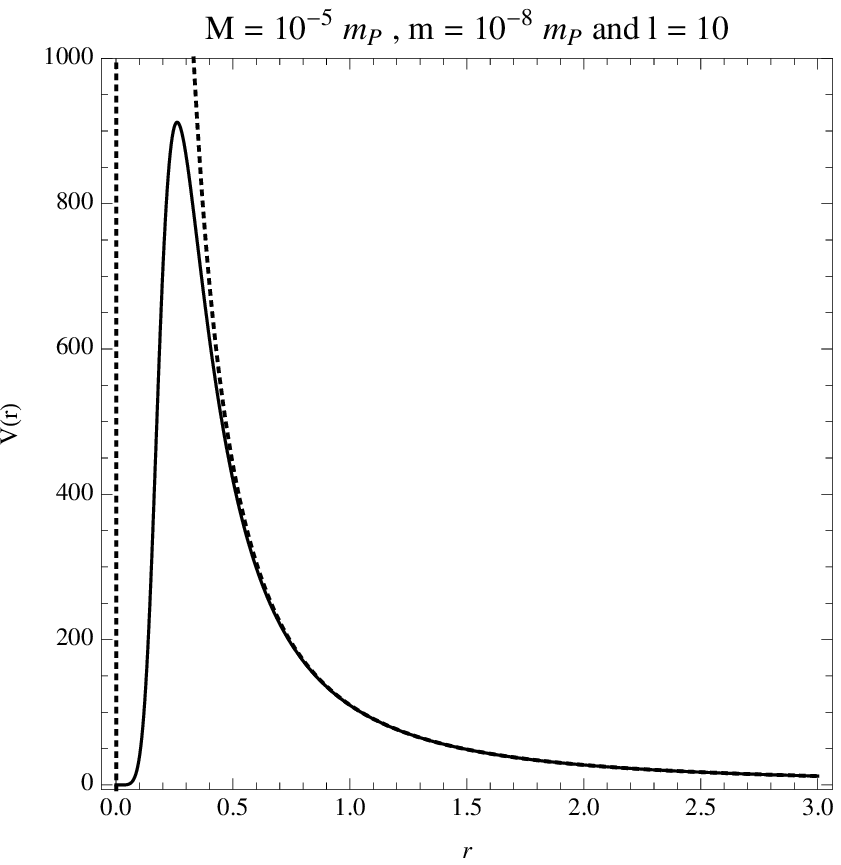}
  \includegraphics[width=6.7cm]{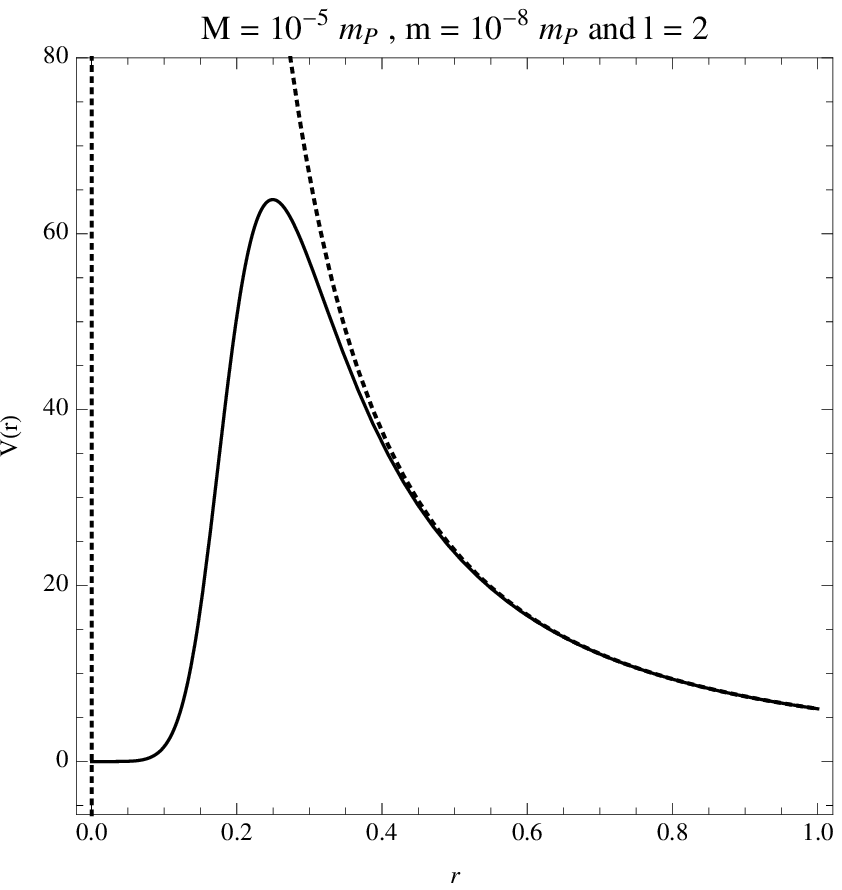}
  \end{center}
  \caption{Effective potential $V(r)$ for the self-dual black hole (solid line) and the classical
black hole (dashed lines) for $l=0,1,2,10$. 
Where $M$ is the black hole mass and $m_{\Phi}$ is the scalar field mass.
}
\label{V}
  \end{figure}

The wave-equation for a massive scalar field in a general curved space-time is
\begin{eqnarray}
\frac{1}{\sqrt{ -g }} \partial_{\mu} \left( g^{\mu \nu} \sqrt{- g} \partial_{\nu} \Phi \right) - m_{\Phi}^2 \Phi =0,
\label{SF}
\end{eqnarray}
where $\Phi \equiv \Phi(r, \theta, \phi, t)$ and $m_{\Phi}$ is the mass of the field. Inserting the metric of the
self-dual black hole we obtain the following differential equation
\begin{eqnarray}
\hspace*{-2.6cm} && 0 = H(r) \left(2  \frac{\partial^2 \Phi }{\partial t^2}
-G(r) F'(r)  \frac{\partial \Phi }{\partial r}     \right)
- 2 G(r) \left(   \frac{\partial^2 \Phi }{\partial  \theta^2}
+\cot  \theta   \frac{\partial \Phi }{\partial  \theta}
+\csc ^2 \theta  \frac{\partial^2 \Phi }{\partial  \phi^2}   \right) \nonumber \\
\hspace*{-2.6cm} &-&F(r) \left(32 m_{\Phi}^2 \csc \theta    \sqrt{\frac{G(r)}{F(r)}}   \Phi
+   H(r) G'(r)  \frac{\partial \Phi }{\partial r}
+2   G(r) H'(r) \frac{\partial \Phi }{\partial r}
+2 G(r) H(r) \frac{\partial^2 \Phi }{\partial r^2} \right) \label{SFGFH}
\end{eqnarray}
where a prime indicates a partial derivative with respect to $r$.
Making use of spherical symmetry and time-translation invariance we write the scalar field as
\begin{eqnarray}
\Phi(r, \theta, \phi, t) := T(t) \, \varphi(r) \, Y(\theta, \phi). 
\label{dec1Phi}
\end{eqnarray}
(The indices $l,m$ on the spherical harmonics $Y_{l,m}$ will be suppressed.)
Using the standard method of separation of variables allows us to split Eq. (\ref{SFGFH})
in three equations, one depending on the $r$ coordinate, one on the
$t$ coordinate and the remaining one depending
on the angular variables $\theta, \phi$.
\begin{eqnarray}
&& \frac{\sqrt{G F}}{H} \frac{\partial}{\partial r}
\left( H \sqrt{G F} \,\,  \frac{ \partial \varphi(r)}{\partial r} \right) - \left[ G \left(m_{\Phi}^2 + \frac{l(l+1)}{H} \right) - \omega^2 \right] \varphi(r)  = 0, \label{radial} \\
&&  \left( \frac{\partial^2  }{\partial  \theta^2}
+\cot  \theta   \frac{\partial  }{\partial  \theta}
+\csc ^2 \theta  \, \frac{\partial^2  }{\partial  \phi^2}   \right) Y(\theta, \phi) = - l(l+1) Y(\theta, \phi),  \\
&& \frac{\partial^2 }{\partial t^2}  T(t) = - \omega^2 T(t).
\label{TpY}
\end{eqnarray}
To further simplify this expression we rewrite it by use of the tortoise coordinate $r^*$ implicitly defined by
\begin{eqnarray}
\frac{d r^*}{d r} := \frac{1}{\sqrt{GF}} ~.
\label{torto}
\end{eqnarray}
Integration yields
\begin{eqnarray}
r^* = r   - \frac{a_0^2}{r \, r_- r_+}
+ a_0^2 \frac{ \left( r_-  +  r_+ \right)}{ r_-^2
   r_+^2 } \log(r)  \nonumber \\
   + \frac{\left( a_0^2 + r_-^4\right)}{r_-^2 (r_- - r_+)}   \log (r - r_-) +
   \frac{\left(a_0^2 + r_+^4\right) }{r_+^2
   (r_+  -  r_-)} \log (r- r_+) ~.
\label{tortoise}
\end{eqnarray}
Further introducing the new radial field $\varphi(r) := \psi(r)/\sqrt{H}$,
the radial equation (\ref{radial}) simplifies to
\begin{eqnarray}
&& \left[\frac{\partial^2}{\partial r^{* 2}} + \omega^2 - V(r(r^*)) \right] \psi(r) = 0 \nonumber \\
&& V(r) = G \left(m_{\Phi}^2 +\frac{ l(l+1) }{H} \right)
+ \frac{1}{2} \sqrt{ \frac{G F}{H}} \left[ \frac{\partial}{\partial r} \left(  \sqrt{ \frac{G F}{H}}  \frac{ \partial H}{\partial r} \right) \right].
\label{simply}
\end{eqnarray}
Inserting the metric of the self-dual black hole one finally obtains the potential to
\begin{eqnarray}
&& V(r) = \frac{(r-r_-) (r-r_+)}{(r^4 + a_0^2)^4} \Big[ \left(a_0^2+r^4\right)^3 m_{\Phi }^2 (r+r_*)^2 \nonumber \\
&&+ r^2 \Big(a_0^4 \left(r
   \left( \, \left(K^2-2\right) r+r_- + r_+\right)+2 K^2 r r_*+K^2 r_*^2 \right) \nonumber \\
   &&  \hspace{-0.5cm}
   +2 a_0^2 r^4
   \left(\left(K^2+5\right) r^2+2 K^2 r r_*+K^2 r_*^2-5 r (r_-+r_+)+5 r_- r_+\right) \nonumber \\ 
&& +r^8
   \left(K^2 (r+r_*)^2+r (r_-+r_+)-2 r_- r_+\right)\Big)  \Big],
\label{VLBH}
\end{eqnarray}
where $K^2=l(l+1)$.
This expression simplifies significantly 
in the S-wave approximation $l=0$,
\begin{eqnarray}
 && V_{00}(r) =  \frac{(r-r_-) (r-r_+)}{\left(a_0^2+r^4\right)^4}
 \Big[\left(   a_0^2+r^4\right)^3 m_{\Phi }^2 (r+r_*)^2 \nonumber \\
&&+ r^2 \Big(a_0^4 r (-2
   r+r_-+r_+) +2 a_0^2 r^4 \left(5 r^2-5 r (r_-+r_+)+5 r_- r_+\right) \nonumber\\ 
&& +r^8 (r
   (r_-+r_+)-2 r_- r_+)  \Big) \Big] ~ .
   \label{VS}
   \end{eqnarray}
The potential is shown for $l=0,1,2,10$ in Figure \ref{V}. The relevant information
we extract from this is that the maximum of the potential is at a radial distance of
$\approx l_{\rm p}/3$ and the potential has
a width of the order $l_{\rm p}$.

Note how very different this behavior is from the usual case. In
the case of small masses $m \ll m_{\rm p}$ that we are interested in, for
the normal Schwarzschild black hole the potential barrier is much closer to
the horizon and much higher than it is here. This new feature is a consequence
of the presence of the second horizon of the solution.

\subsection{Horizon to Pinhole Distance}

With this preparation we are now in the position to examine the relevance of the potential
wall for the parameter ranges we are studying here. We will see that the greybody factors
will be negligible for an interesting reason: the distance between the peak of the potential
wall and the horizon is always many orders of magnitude smaller than the wavelength of the
emitted particle. Recalling that the black hole's particle emission is a tunneling process
through the horizon, the additional potential barrier is simply trespassed together with
the horizon and does not influence the emission spectrum in the mass- and wavelength-regime
we are interested in. Another way to say this is that with the typical wavelengths we
are considering here, the emitted particles are not localized enough so they can even
be considered emitted to within the potential wall. 

To see this,
let $l(m)$ be the physical distance between the horizon and the pinhole for an ultra-light 
black hole of mass $m$ in the small $\delta$ limit
\begin{eqnarray}
l(m) &=& \int_{r_+}^{\sqrt{a_0}} \sqrt{g_{rr}}dr 
= \int_{r_+}^{\sqrt{a_0}} \sqrt{\frac{r^4+a_0^2}{(r-r_+) r^3}} dr~.
\eeqn
We see that
\beqn
l(m)/\sqrt{2}\leq \int_{r_+}^{\sqrt{a_0}} \frac{a_0}{\sqrt{(r-r_+) r^3}}dr &:=& j(m) \leq l(m)~.
\end{eqnarray}
For $4m=2r_+\leq \sqrt{a_0}$, which is always fulfilled for our case, we have
\begin{eqnarray}
j(m) &=& \int_{0}^{\sqrt{a_0}-r_+} \frac{a_0}{\sqrt{x (x+r_+)^3}}dx 
\nonumber \\
&=&  \int_{0}^{r_+} \frac{a_0}{\sqrt{x (x+r_+)^3}}dx + \int_{r_+}^{\sqrt{a_0}-r_+} \frac{a_0}{\sqrt{x (x+r_+)^3}}dx~,
\eeqn
and
\beqn
j(m)/\sqrt{2}\leq \int_{0}^{r_+} \frac{a_0}{\sqrt{x (r_+)^3}}dx +  \int_{r_+}^{\sqrt{a_0}-r_+} \frac{a_0}{(x+r_+)^2}\leq 2^{3/2}j(m)~.
\end{eqnarray}
Integrating and combining the previous inequalities we find that
\begin{eqnarray}
l(m)/2\leq \frac{5}{4}\frac{a_0}{m} - \sqrt{a_0}\leq 2^{3/2}l(m) ~.
\end{eqnarray}
This quantity $l(m)$ now has to be compared to the inverse of the temperature given by 
Eq. (\ref{Temperatura}). In the limit of $m \ll m_{\rm p}$ one finds, after re-inserting
the Planck mass,
\beqn
l(m) < \frac{5}{4} \frac{1}{m} \ll \frac{\pi}{2} \frac{1}{m} \left( \frac{m_{\rm p}}{m} \right)^2 \approx \frac{1}{T(m)} ~.  
\eeqn

\begin{figure}
 \begin{center}
 \hspace{-0.3cm}
  \includegraphics[height=7.0cm]{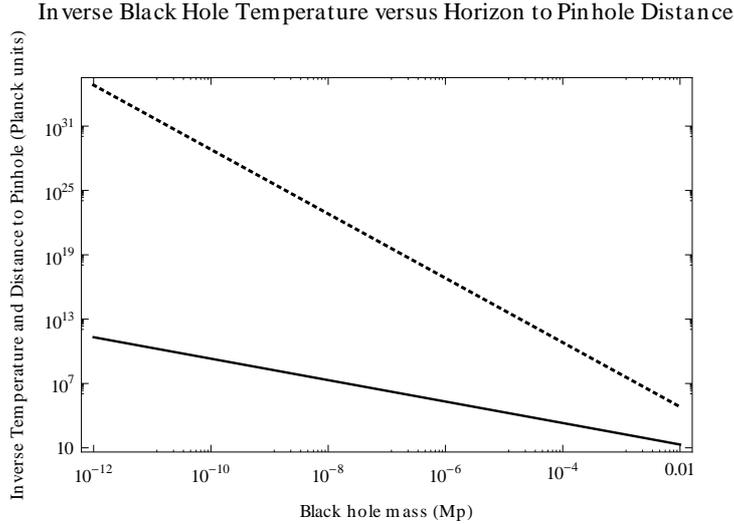}
      \end{center}
  \caption{\label{Pot}
  $1/T(m)$ (dashed) and $l(m)$ (solid) in units of the Planck mass as a function of the black hole's mass $m$.}
\label{disttopin}
  \end{figure}

For a visual comparision, the quantities $l(m)$ and $1/T(m)$ are also plotted in Figure \ref{Pot}. One sees that, in
the limit of black hole masses much smaller than the Planck mass, the inverse temperature,
or the average wavelength of the emitted particles, is always many orders of magnitude larger
than the distance between the horizon and the pinhole. 
The numerical investigation also confirms that for $m<10^{-2}$ we have to good accuracy $l(m)\approx 1/m$.

To summarize this and the previous section, we conclude that in limit we are interested in
we can neglect the details of the potential wall and use the usual low energy approximation.
In this
limit the greybody factors take the values \cite{Page:1976df}
\beqn
\Gamma_{1/2}(\omega,M) = \frac{\pi}{2} l_{\rm Pl} ~,
\eeqn
for spin $1/2$ and
\beqn
\Gamma_{1}(\omega,M) =\frac{4 \pi}{3} l_{\rm Pl} (\omega l_{\rm Pl}) ~,
\eeqn
for spin 1, where we have taken into account that the effective surface of the black hole is given by the
size of the pinhole rather than the size of the horizon.

\subsection{Particle Flux}
\label{particle}

With the greybody factor $\Gamma_s(\omega,M)$ and the number-density of the emitted radiation from the previous
sections, the number of particles of type $j$ emitted in the energy range
between $\omega$ and $\omega + d \omega$ per time is now
\begin{eqnarray}
\frac{ d^2 N_j}{d \omega d t} = \frac{1}{2\pi} g_j \Gamma_s (\omega,M) n_j(\omega,M) \quad.
\label{rate}
\end{eqnarray}

Where $g_j$ is the number of degrees of freedom of the particle and $n_j$ depends on
the mass of the particle and its spin, though both can be neglected for the energy
range we are interested in. Since most elementary
particles are unstable, we further have to take into account the decay channels of the
primary particles. We denote the number of particles of type $X$ with energy $E$
produced by the parent $j$ with energy $\omega$ as
\beqn
\frac{d f_{j X} (\omega, E)}{dE} ~.
\eeqn
It is then
\begin{eqnarray}
&&\frac{df_{j j} (\omega, E)}{dE} = \delta(\omega - E)~,  \\
&& \int  \frac{df_{j X} (\omega, E)}{dE} dE = N(j \rightarrow X)~,
\end{eqnarray}
where $N(j \rightarrow X)$ is total number of particles of type $X$ produced by the
initial particles $j$. With that parameterization the flux of particles of type
$X$ from the hole is 
\begin{eqnarray}
 \frac{ d^2 N_X}{d E d t} = \frac{1}{2 \pi} \sum_j g_j \int_{\omega = E}^{\omega = M} \Gamma_j(\omega, M)  n_j(\omega,M) \frac{df_{j X} (\omega, E)}{dE} d\omega~.
\label{rate2}
\end{eqnarray}
This flux is that of a single black hole in rest. If the
black hole emits a particle, it will aquire a recoil into the opposite direction of the particle.
In our case however, due to the low temperature of the black holes, the average velocity that
the black hole aquires in this process is $T(m)/m \ll 1$. The typical velocities of the black holes in
a dark matter halo thus have a negligible influence on the emission spectrum. Note again how very
different this is to the usual case of Schwarzschild black holes in the final stages of evaporation. 

To obtain the particle flux received on Earth, we consider a dark matter halo of mass $M_{\rm DM}$ at a redshift $z$. 
From the total mass, one obtains the approximate number of black holes it contains. One further has to 
take into account the drop of luminosity with distance. With the fragmentation into protons from Eq. (\ref{rate2}), 
this yields for example for the number of protons with energy $E$ detected on Earth per
time and unit area $A$
\beqn
\frac{ d^3 {\cal N}_p}{d E d t d A} = \frac{M_{\rm DM}}{m} \frac{1}{4 \pi r_{\rm cm}(z)^2(1+z)^2} \frac{ d^2 N_p}{d E d t} ~.
\label{totnum}
\eeqn
Here, $r_{\rm cm}(z)$ is the comoving distance (line-of-sight) to the object and, for standard $\Lambda$CDM--model ($k$=0) , given by
\beqn
r_{\rm cm} = \frac{1}{H_0} \int_0^z \frac{dz}{\sqrt{\Omega_M (1+z)^3 + \Omega_\Lambda}}~,
\eeqn
where $H_0, \Omega_M, \Omega_\Lambda$ are the standard values for $\Lambda${\sc CDM} cosmology.
One would expect
that the black holes in the dark matter halo do not all have exactly the same mass, but that the
distribution is smeared out over some range of masses, and thus Eq. (\ref{totnum}) has to be averaged over this
mass distribution. 

\section{Conclusion}
\label{further}

We have derived here an approximate analytic expression for the emission spectrum of
self-dual black holes in the mass and temperature limits valid for primordial black holes evaporating today. 
The idea that primordial black holes are dark matter candidates is appealing since it is very minimalistic
and conservative, requiring no additional, so far unobserved, matter. This idea has therefore received
a lot of attention in the literature. However, the final 
stages of the black hole evaporation seem to be amiss in observation, and so there is a need to explain
why primordial black holes were not formed at initial masses that we would see 
evaporating today. The self-dual black holes we have studied here offer a natural explanation since
they evaporate very slowly. The analysis we have presented here allows to calculate the
particle flux from such dark matter constituted of self-dual black holes, and therefore is instrumental
to test the viability of this hypothesis of dark matter constituted of self-dual black holes 
against data.

\section*{Acknowledgements}

SH thanks the Perimeter Institute for Theoretical Physics for hospitality during
the work on this manuscript.
We thank Alberto Montina for the assistance given in a difficult calculation,
and Stefan Scherer and Andr\'e Yoon for helpful conversation. 
Research at
Perimeter Institute is supported by the Government of Canada through Industry Canada
and by the Province of Ontario through the Ministry of Research \& Innovation.

\end{document}